\documentclass[preprint,showpacs,preprintnumbers,amsmath,amssymb]{revtex4}
\usepackage{graphicx}

\begin{document}
\title{Phonon in SrTiO$_3$ under finite electric fields}
\author{Ivan I. Naumov and Huaxiang Fu}
\affiliation{Department of Physics, University of Arkansas,
Fayetteville, Arkansas 72701, USA}

\date{\today}

\begin{abstract}
We propose an efficient approach within the density-functional
theory to determine the phonon structure of infinite solids under
finite electric fields. We apply this approach to technological
SrTiO$_3$, predicting many unusual properties---including a giant
frequency shift, anomalous piezoelectric response, as well as
striking dielectric tunability. A microscopic understanding for
individual phonon modes under finite electric fields is also
provided.
\end{abstract}

\maketitle

It has been nearly 40 years since Worlock and Fleury discovered by
means of Raman scattering that {\it under finite electric fields}
the transverse-optical (TO) phonon in SrTiO$_3$,\cite{Fleury1} as
well as in KTaO$_3$,\cite{Fleury2} exhibits a striking shift in
frequency---by more than 400\% of its zero-field value---when a
very small field of 12 kV/cm is applied.\cite{Fleury1} Despite its
obvious importance, this giant field-induced shift of phonon
frequency has not thus far been independently investigated and/or
understood via the powerful density-functional theory (DFT) which,
on the other hand, has been amply applied in the past three
decades on a variety of {\it other} material properties. The
reason lies in the difficulty in handling finite electric fields
within the direct DFT approach, a topic that has recently drawn a
great deal of interest and
progress.\cite{DV1,Umari,Sai,Nunes,Fu1,DV2,Kunc}

SrTiO$_3$ is an incipient ferroelectric that is approaching, but
does not in fact undergo, a paraelectric-to-ferroelectric phase
transformation when lowering temperature, and is often considered
as an ``inert'' perovskite.\cite{Lines} However, being on the edge
of phonon softening, SrTiO$_3$ shows unusual sensitivity on
external electric fields that are able to drive easily the
occurrence of the phase transition, and thus likely lead to an
anomaly in the field-induced structural response. Given their
technological importance, this and other similar ferroelectric
incipients (in particular, when they are under finite electric
fields) remain, nevertheless, poorly understood. For instance, it
is not clear how different phonon modes may respond differently to
the electric fields in terms of their frequencies and in terms of
their atomic participations (that is, the phonon properties).
Furthermore, little is known on how atomic displacements driven by
external fields will couple with strain in incipients (i.e., the
mechanical properties), and whether SrTiO$_3$ in {\it finite}
electric fields can be a good piezoelectric, in addition to being
a good dielectric. Finally, a direct DFT modelling of the
``dielectric tunability'' by finite electric fields in incipients
(i.e., the dielectric properties) is still at its early stage and
thus much needed.\cite{DV-APS}

The purpose of this Letter is threefold: (1) to show that it is
possible to devise a direct density-functional approach to
determine and understand the phonon spectrum of infinite solids
under finite electric fields; (2) to demonstrate that the approach
works robust for incipient SrTiO$_3$, and in particular, the
theoretical prediction of field-induced frequency shift is in
excellent agreement with the result observed in experiments; (3)
to present other unusual properties in SrTiO$_3$ that occur
``only'' under finite electric fields, that is, a gigantic
piezoelectric response as well as a superior field-induced
dielectric tunability.

To determine phonon structure, namely vibration frequencies and
phonon eigenmodes, of a solid under finite electric fields, we
start with free energy ${\rm F}={\rm U}-{\bf E}\cdot {\bf P}$,
where {\bf E} (assumed here to be along the pseudo-cubic [001]
direction) and {\bf P} are the screened macroscopic field and
polarization per cell, respectively. This yields force-constant
matrix elements as
\begin{eqnarray}
\frac{\partial ^2F}{\partial r_{i\alpha } \partial r_{j\beta }} =
\frac{\partial ^2U}{\partial r_{i\alpha } \partial r_{j\beta }} -
E\frac{\partial Z^*_{3\alpha }(i)}{\partial r_{j\beta }} ,
\label{Edyn}
\end{eqnarray}
where $r_{i\alpha }$ is the position of atom {\it i} along the
$\alpha $ cartesian direction, and $Z^*(i)$ is the Born effective
charge tensor of this atom. It should be emphasized that each term
in the right-hand side of Eq.(\ref{Edyn}) must be evaluated at
{\it field-induced} new atomic-positions \{{\bf R}$_{\rm new}$\}
and cell-shape strain \{$\eta _{\rm new}$\}, instead of the
zero-field \{{\bf R}$_{\rm 0}$\} and \{$\eta _{\rm 0}$\}.

Determination of the force-constant matrix of Eq.(\ref{Edyn}) is
done in two steps. First, \{{\bf R}$_{\rm new}$\} and \{$\eta
_{\rm new}$\} are obtained using a newly-developed method that was
described in Ref.~\onlinecite{Fu2}. Second, the force-constant
matrix is evaluated as follow. The first term in Eq.(\ref{Edyn}),
which is associated with the {\it zero-field} internal energy {\it
U} but at displaced atomic positions, is ready to be computed
using standard density-functional perturbation theory
(DFPT).\cite{Gonze,Krakauer} The second term requires, however, a
perturbation theory of the macroscopic polarization {\bf P} with
respect to the atomic displacements, and is rather cumbersome to
deal within the DFPT. Here, we numerically find (by performing
finite-difference calculations) that, even for an electric field
as large as 0.5$\times 10^8$ V/m, the second term in
Eq.(\ref{Edyn}) turns out to be less than 0.5\% of the first term
for all diagonal elements, and thus {\it contributes practically
very little to the dynamic matrix}. This particular finding allows
us to neglect the insignificant second term. The resulting
simplicity and efficiency in the present approach is precisely
what we pursue, and allows the determination of the phonon
structure under finite electric fields to be routinely performed
within the DFT. This approach works in general for phonon at any
wave-vector {\bf q}, whereas in the following we focus on the
phonon modes at the zone center since they can be directly
compared with Raman experiments.\cite{Fleury1,Fleury2}

It should be pointed out that, apart from the (externally-applied)
field {\bf E}, a long-wavelength longitudinal-optical (LO) phonon
in polar materials will generate a second macroscopic field {\bf
E$_1$} with its direction depending on the phonon wave-vector {\bf
q}. Here we choose to study a configuration that likely leads to
most pronounced effects that are to be caused by the applied field
{\bf E}, namely, when {\bf E} and {\bf q} are {\it perpendicular},
with {\bf q} chosen along the pseudo-cubic [010] direction.

Obviously, quantum fluctuation effects in SrTiO$_3$ at low
temperature, which are responsible for its incipient behavior, can
not be described by zero-temperature DFT calculations and are not
the focus of this work. On the other hand, the DFT is able to
predict quantitatively the electric-field effects in SrTiO$_3$,
since these effects are mainly determined by the zero-field
frequency (i.e., the energy surface) of the ongoing soft mode
(assuming that this frequency is correctly reproduced by DFT).
Furthermore, anti-ferroelectric rotation in low-temperature
SrTiO$_3$ is not considered here, as it affects mainly the
zone-boundary (not the zone-center) phonons.

Technically, our calculations are carried out using the ABINIT
code,\cite{ABINIT1} modified to perform {\it finite-field}
geometry optimization. A sufficient energy cutoff of 110 Ry is
used to ensure a satisfactory convergence. The zero-field lowest
TO phonon, that turns soft in {\it normal} ferroelectrics, is
found to be still stable and at the frequency of 13.2 cm$^{-1}$
for cubic SrTiO$_3$ with the lattice constant 3.892 {\AA}. This
frequency is in good agreement with experimental value of 11
cm$^{-1}$\cite{Fleury1}, and in fact, an agreement is found for
all modes across the phonon spectrum.

Fig.\ref{Fphn} shows the phonon spectrum under an electric field
of 0.22$\times 10^8$ V/m, compared to the zero-field case. The
most striking results are: (1) Upon the electric field, the
degeneracy is completely lifted for all phonon modes, and in
particular, for the zero-field ``silent'' modes (being
triply-degenerate with $\Gamma _{25}$ symmetry in cubic
SrTiO$_3$). Being {\it non-polar} with only oxygen amplitude,
these silent modes were commonly thought to split into one singlet
and one doublet by electric fields \cite{Fleury_PRB}. (2) The
lowest two modes in the lower panel of Fig.\ref{Fphn}, which split
from a doublet TO phonon in the upper panel, undergo {\it giant}
shifts in frequency from 13.2 cm$^{-1}$ at zero field to 48.7 and
92.2 cm$^{-1}$ at the considered magnitude of field. In other
words, the frequencies of these modes are altered by the applied
electric field by more than 400\% and 800\% (!), respectively. (3)
Another mode at zero-field frequency of 163.2 cm$^{-1}$ (notice
that this is {\it not} the ongoing soft mode) displays also a
remarkable frequency shift by as much as 26.6 cm$^{-1}$. (4) On
the other hand, modes at 447.0 and 772.4 cm$^{-1}$ are found to be
{\it least} affected (if any).

In order to understand the {\it microscopic} origin of why some
modes, not others, are drastically altered by electric fields, and
furthermore, to provide valuable guide for understanding future
experimental results, Table \ref{Tmode} gives the vibration
eigenvectors of the phonon modes. Surprisingly, the second mode in
this Table---namely, the mode displaying the largest field-induced
frequency shift, vibrating in parallel with the {\bf E} direction,
and its frequency to be referred to as $\omega _{||}$---turns out
to have {\it virtually no} amplitude from Ti, instead it is
predominantly from strontium and oxygen atoms. This is in contrast
with the zero-field situation where Ti contribution is large
($\sim $ 0.31) for the corresponding mode. On the other hand,
another (also {\it z}-axis polarized) mode at 189.8 cm$^{-1}$
shows an opposite trait, with a {\it heavy} Ti involvement. Also
interestingly, it can be seen in Table \ref{Tmode} that a large
field-induced frequency shift occurs not only to those modes that
vibrate {\it parallel} to {\bf E}, as it should (for instance, see
the modes at 92.2, 189.8, and 549.9 cm$^{-1}$)---and in fact, the
applied field also produces a remarkable shift to the mode that is
{\it transverse} to the field direction (see, the first mode in
Table \ref{Tmode}, and the frequency of this mode will be denoted
as $\omega _{\perp }$ hereafter). Meanwhile, the modes that
exhibit the least frequency shift (e.g., modes at 447.0 cm$^{-1}$
and at 772.4 cm$^{-1}$) are found to associate with vibrations
along the [010] direction. This can be explained by the fact that
these modes originate from the {\bf E}$_1$ field due to the LO/TO
splitting and thus subject to little effects from the applied {\bf
E} field.

We now wonder how the phonon frequencies may vary with different
field-strengths, and the results are depicted in Fig.\ref{Fedep}a
for the $\omega _{||}$ and $\omega _{\perp }$ modes. Intriguingly,
our results show that the field-induced increase in frequency
(i.e., so-called ``phonon hardening'') is characterized by two
drastically different regimes, separated by a critical field E$_c$
and equal to $\sim $0.5$\times 10^8$ V/m. Below E$_c$, $\omega
_{||}$ and $\omega _{\perp}$ increase {\it dramatically} and
strong {\it nonlinearly}---with their difference $\omega
_{||}-\omega _{\perp }$ first enlarged and then saturated at
E$_c$---as the field strength increases. On the other hand and
above E$_c$, $\omega _{||}$ and $\omega _{\perp}$ show almost a
{\it linear} field-dependence, and interestingly, the
frequency-vs-field slopes are nearly equal for these two modes,
whereas they vibrate along completely different directions.
Fitting the two curves in Fig.\ref{Fedep}a at high field-strength
($E>0.25\times 10^8$ V/m) to an analytical expression $\omega
^2(E)+\omega ^2(E=0)=\sigma E^\gamma $ (derived from the
Ginzburg-Landau-Devonshire theory\cite{Fleury1}) yields, for the
$\omega _{||}$ and $\omega _{\perp}$ modes, an exponent $\gamma $
of 0.31 and 0.56, respectively. Our direct DFT results thus
demonstrate that the exponent $\gamma $ not only significantly
varies from mode to mode, but also considerably deviates from the
value of 0.66 from the phenomenological theory.\cite{Scale}

The peculiar behavior on the field-dependence of the mode
frequency in Fig.\ref{Fedep}a is found to arise from the atomic
displacements and cell-shape changes induced by the electric
fields, shown in Fig.\ref{Fedep}b. It can be clearly seen that
below the critical E$_c$ field the dramatic and nonlinear
displacements of O atoms (with respect to Ti), as well as the
sharp increase of the c/a strain, correlate well with the phonon
responses in Fig.\ref{Fedep}a. Notably, two {\it non}-equivalent
oxygen atoms (namely, O$_1$ and O$_3$) respond remarkably similar
to the applied electric field, whereas the (relative) displacement
of Sr is almost negligible. These results thus show (not
surprisingly) that the field effect is primarily to move Ti with
respect to the rather rigid oxygen octahedral, while Sr
contributes by having the ``right'' atomic size to form an
incipient (Indeed, the low-frequency modes in Table \ref{Tmode},
which matter most for the incipient behavior, all show a
significant Sr participation).

Our theoretical attempt on phonon under finite electric fields
yields results that quantitatively agree with experiments. For
example, besides a sharp Raman peak that results from the $\omega
_{||}$ phonon, one additional peak, despite its weak intensity,
was clearly observed at a lower frequency.\cite{Fleury1} This
additional peak is consistent with the existence of the $\omega
_{\perp}$ mode as predicted in our calculations. Furthermore, our
theoretical $\omega _{||}$ frequency at the 0.012$\times 10^8$ V/m
field is determined to be 42.7 cm$^{-1}$, in excellent agreement
with the experimental value of 45 cm$^{-1}$, measured under the
same field strength and at temperature 8\,K.\cite{Fleury1}

Finally, we would like to examine other technological
properties---namely, piezoelectric $e_{33}-e_{31}$ coefficient
(which measures the piezoelectric response to a tetragonal strain
when volume is preserved) and {\it static} dielectric $\varepsilon
_{33}$ and $\varepsilon _{11}$ constants---of SrTiO$_3$ that is
placed under {\it finite electric fields}. The $e_{33}-e_{31}$
coefficient is null at zero field (see, Fig.\ref{Fedep}a). This
coefficient becomes, however, strikingly large---being as much as
19 C/m$^2$ under a small field of 0.012$\times 10^8$ V/m. Note
that this magnitude of piezoelectric response in SrTiO$_3$ (which
is predicted here for the first time, to our knowledge) exceeds
far more than the value of 4.08 C/m$^2$ in PbTiO$_3$\cite{Cohen},
and even than the intrinsic theoretical $e_{33}$ value of 9.1
C/m$^2$ in PMN-PT~\cite{LB1}. We attribute this exceptional
response to the field-induced occurrence of the
paraelectric-to-ferroelectric phase transition in SrTiO$_3$, which
does not happen in PbTiO$_3$ and in PMN-PT. Interestingly, the
$e_{33}-e_{31}$ coefficient in Fig.\ref{Fedep}a first increases,
then decreases, and eventually levels off at a nonzero value of
$\sim $ 6 C/m$^2$. Regarding the dielectric properties, the
$\varepsilon _{33}$ dielectric constant is determined to be 12215,
1120, 188, 117, 77, and 61 for the six considered field strengths
at 0, 0.012, 0.22, 0.50, 1.0, and 1.5 $\times 10^8$ V/m,
respectively. This tells us that the $\varepsilon _{33}$ constant
decreases dramatically by more than {\it one order of magnitude}
when going from zero field to a very small field of 0.012$\times
10^8$ V/m, demonstrating the existence of an incredible
field-induced dielectric tunability! This giant tunability is
found also true for the $\varepsilon _{11}$ component, being
12215, 7123, 866, 485, 337, 264 at the field strengths given
above. Last (but not least), we want to point out that our direct
DFT calculations of phonon frequencies and ({\it independently})
dielectric constants prove that the Lyddane-Sachs-Teller (LST)
relationship \cite{LST} is fully valid under finite electric
fields. In fact, the {\it static} dielectric constants obtained
from direct DFPT calculations and those from the LST relationship
(using phonon frequencies) are found to differ by less than 0.1\%.

In summary, a practical and efficient DFT approach that allows to
determine the phonon structure of solids in finite electric fields
was developed, and applied, to study the electric-field effects on
the zone-center modes of incipient SrTiO$_3$, yielding a good
agreement with experiment. Our results confirm quantitatively the
striking field-induced frequency shifts that have been
experimentally observed---but not explained by direct
first-principles theory---for nearly four decades. Our work
further provides a microscopic understanding of individual phonon
mode under finite fields.  We also find: (i) A giant phonon
hardening (as the frequency increases by {\it several fold} with
respect to the zero-field value) takes place, not merely for the
$\omega _{||}$ mode that vibrates in parallel with---but also for
the $\omega _{\perp }$ mode that, on the other hand, vibrates
perpendicularly to---the direction of the applied electric field.
(ii) After hardening, the mode showing the largest frequency shift
(i.e., the $\omega _{||}$ mode) has a surprisingly small Ti-atom
participation. (iii) The mode at the zero-field frequency of 163.2
cm$^{-1}$, whereas {\it not} being the ongoing soft mode, displays
also a remarkable response to the imposed electric field. Modes at
447 and 772 cm$^{-1}$ are found to be {\it least} affected by the
{\bf E} field. (iv) The field dependence of the soft-mode
frequency exhibits a crossover at E$_c$=0.5$\times 10^8$
V/m---below which $\omega _{||}$ and $\omega _{\perp }$ upshift
sharply and nonlinearly, but differently. Above E$_c$, the
responses of these two modes are interestingly similar with a
nearly equal frequency-vs-field slope. This dependence is found to
result from the field effects on atomic movements and on the
tetragonal c/a strain. (v) A small electric field of 12 kV/cm is
able to decrease the $\varepsilon _{33}$ constant by one order of
magnitude, showing an extraordinary dielectric tunability. (vi) An
unusual piezoelectric response is predicted to exist in SrTiO$_3$
when placed under finite fields. We expect that the present
approach will have widespread applications for determining phonon
structure and dielectric tunability in solids under finite
electric fields.

We gratefully thank David Vanderbilt for many stimulating
suggestions and discussions. This work was partially supported by
the Office of Naval Research (N00014-03-1-0598). The computing
facility was provided by the Center for Piezoelectrics by Design
(N00014-01-1-0365) as well as by the National Science Foundation
(DMR-0116315).

\newpage
\begin{figure}
\caption{Mode frequency and degeneracy of SrTiO$_3$ under zero
field (the upper panel) and under a field of 0.22$\times 10^8$ V/m
(the lower panel). Zero-frequency acoustic modes are not shown.
Dotted lines indicate the evolution of selected phonon modes.}
\label{Fphn}
\end{figure}

\begin{figure}
\caption{Dependence of (a) the $\omega _{||}$ and $\omega
_{\perp}$ frequencies, and (b) the {\it z}-direction displacements
of Sr, O$_1$ and O$_3$ atoms (using the left vertical-axis) and
the c/a strain (using the right vertical-axis), as a function of
field strength. Inset in part (a) is the piezoelectric
$e_{33}-e_{31}$ coefficient. In part (b), Ti is chosen as the
reference for measuring the {\it z}-axis displacements of other
atoms. Symbols are calculation results; lines are guide for eyes.}
\label{Fedep}
\end{figure}

\begin{table}
\caption{Frequency $\omega $ and mode eigenvector $\xi _\tau $ of
the zone-center phonons in SrTiO$_3$ under a field of 0.22$\times
10^8$ V/m, as well as frequency shift $\Delta \omega =\omega
(E)-\omega (E=0)$. One eigenvector has three vibration components
$\xi _x$, $\xi _y$ and $\xi _z$ (as specified by index $\tau $);
each component is described using the notation $\xi _\tau =(u_{\rm
Sr}, u_{\rm Ti}, u_{\rm O_1}, u_{\rm O_2}, u_{\rm O_3})$ with five
amplitudes from Sr, Ti, O$_1$, O$_2$, and O$_3$ atoms,
respectively. O$_3$ is located beneath the Ti atom. Those
components not given in the table are zero. The {\it x}, {\it y},
and {\it z}-axes are, respectively, pointing along the
pseudo-cubic [100], [010], and [001] directions. }
\begin{tabular}{lll}
\hline \hline
$\omega $ (cm$^{-1}$) &  $\Delta \omega $ (cm$^{-1}$)  &  eigenvectors  \\
\hline
48.7  &  35.5  & $\xi _x$=( 0.43,  0.27, -0.43, -0.53, -0.52) \\
92.2  &  79.0  & $\xi _z$=(-0.57, -0.04,  0.51,  0.51,  0.40) \\
149.9 &  1.2   & $\xi _y$=(-0.71,  0.60,  0.23,  0.16,  0.24) \\
166.3 &  3.1   & $\xi _x$=(-0.58,  0.82, -0.02, -0.03,  0.01) \\
189.8 &  26.6  & $\xi _z$=(-0.44,  0.86, -0.17, -0.17, -0.11) \\
226.0 &  2.2   & $\xi _y$=(-0.01,  0.01,  0.71,  0.00, -0.71) \\
226.4 &  2.6   & $\xi _x$=(-0.01,  0.02,  0.00,  0.70, -0.71) \\
227.0 &  3.2   & $\xi _z$=( 0.00,  0.00, -0.71,  0.71,  0.00) \\
447.0 &  0.0   & $\xi _y$=(-0.05, -0.43,  0.59, -0.34,  0.59) \\
542.3 &  1.2   & $\xi _x$=( 0.04,  0.02, -0.85,  0.37,  0.37) \\
549.9 &  8.8   & $\xi _z$=( 0.03,  0.04,  0.36,  0.36, -0.86) \\
772.4 &  0.6   & $\xi _y$=(-0.12, -0.45,  0.08,  0.88,  0.08) \\
\hline \hline
\end{tabular}
\label{Tmode}
\end{table}

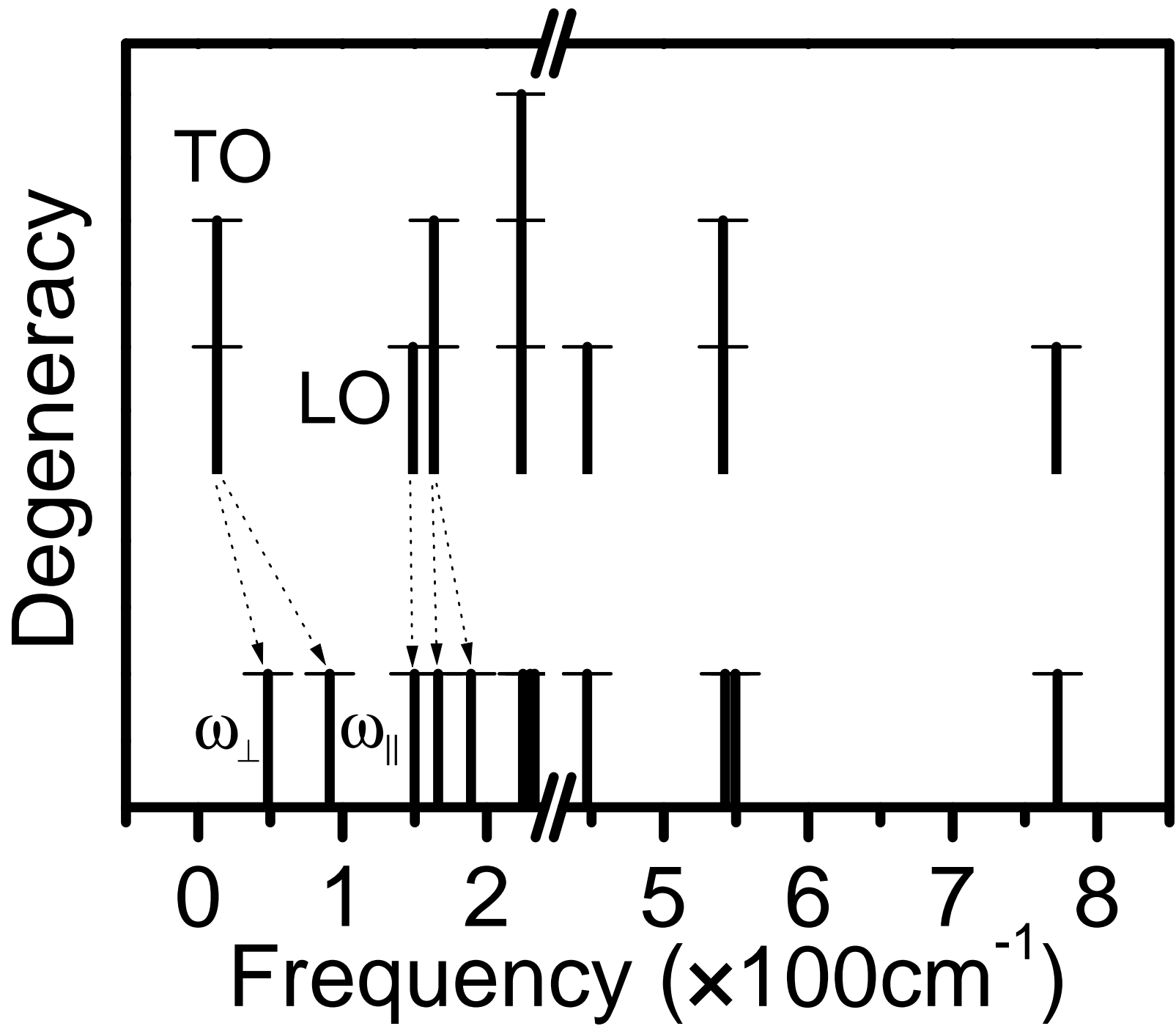
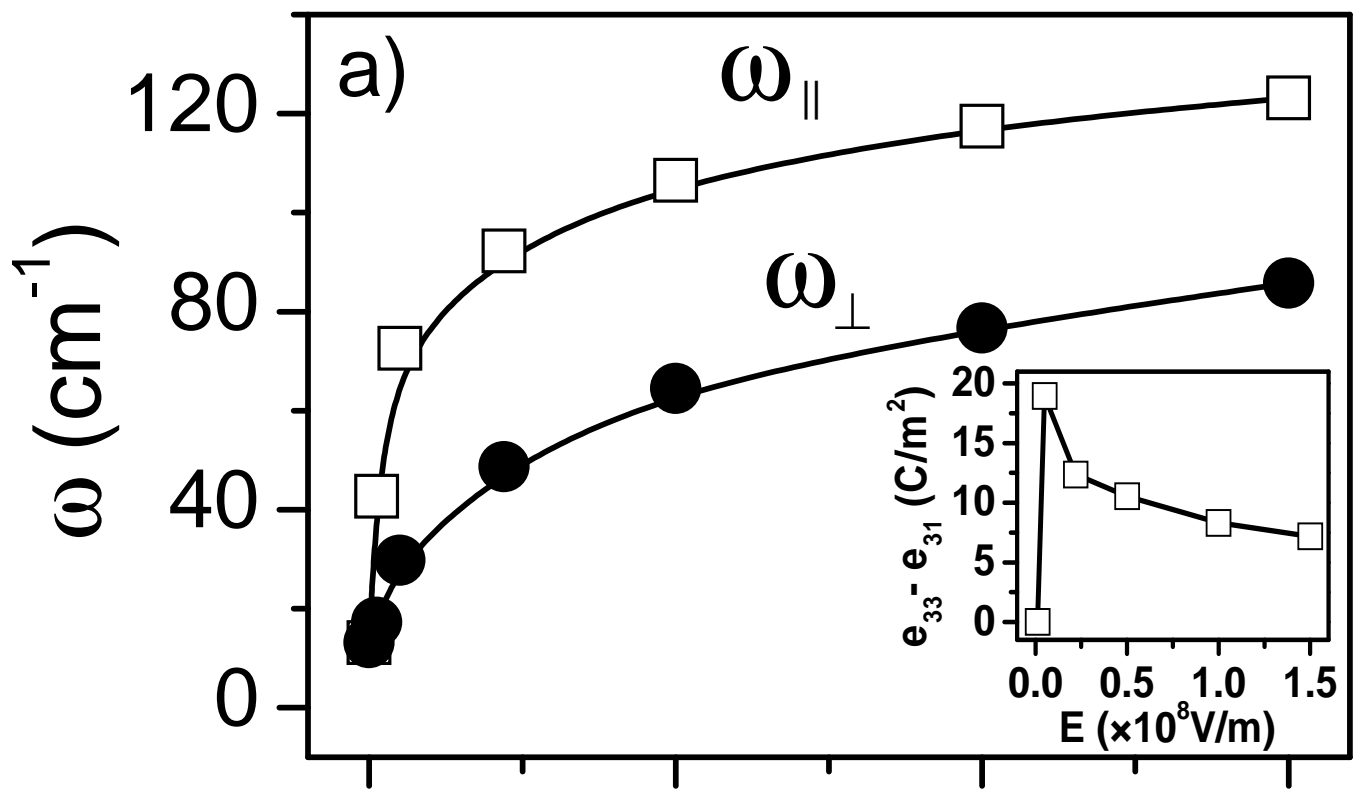
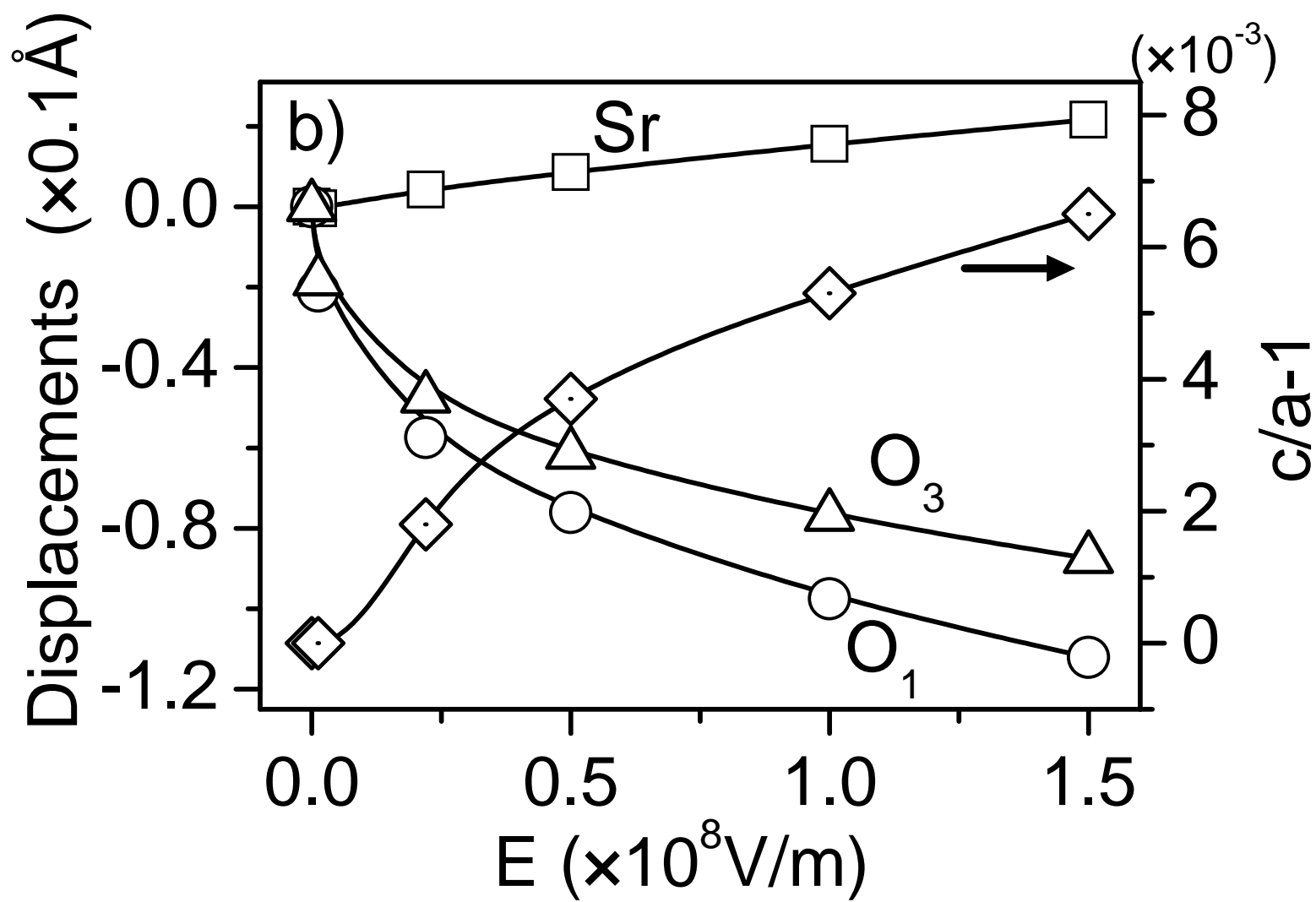
\end{document}